\documentclass[onecolumn]{elsart3p}

\usepackage{graphics}
\usepackage{graphicx}
\usepackage{dcolumn}

\newcommand{\ar}{\arrowvert}
\newcommand{\ra}{\rangle}
\newcommand{\la}{\langle}

\newcommand{\be}{\begin{equation}}
\newcommand{\ee}{\end{equation}}
\newcommand{\ba}{\begin{eqnarray}}
\newcommand{\ea}{\end{eqnarray}}

\begin{document}

\begin{frontmatter}



\title{\vspace{-3.2cm}
\hfill {\scriptsize FZJ-IKP-TH-2008-19, HISKP-TH-08-25}\\
\vspace{5mm}%
Quark mass dependence of the pion vector form factor}


\author{Feng-Kun Guo$^{1}$, Christoph Hanhart$^{1,2}$,
Felipe J. Llanes-Estrada$^3$, Ulf-G. Mei{\ss}ner$^{1,2,4}$}
\address{$^1$ Institut f\"ur Kernphysik and J\"ulich Center for Hadron Physics, Forschungszentrum J\"ulich, D--52425 J\"{u}lich, Germany \\
$^2$ Institut for Advanced Simulations, Forschungszentrum J\"ulich, D--52425 J\"{u}lich, Germany \\
$3$ Departamento de F\'{\i}sica Te\'orica I, Universidad
Complutense de Madrid, 28040 Madrid, Spain \\
$^4$ Helmholtz-Institut f\"ur Strahlen- und
        Kernphysik (Theorie) and Bethe Center for Theoretical Physics,\\ Universit\"at Bonn, D--53115 Bonn, Germany \\
}

\begin{abstract}
  We examine the quark mass dependence of the pion vector form factor,
  particularly the curvature (mean quartic radius). We focus our study
  on the consequences of assuming that the coupling constant of the
  $\rho$ to pions, $g_{\rho\pi\pi}$, is largely independent of the quark
  mass while the quark mass dependence of the $\rho$--mass is given by
  recent lattice data. By employing the Omn\`es representation we can provide a
  very clean estimate for a certain combination of the curvature and
  the square radius, whose quark mass dependence could be determined
  from lattice computations. This study provides an independent access
  to the quark mass dependence of the $\rho \pi \pi$ coupling and in
  this way a non-trivial check of the systematics of chiral
  extrapolations.  We also provide an improved value
  for the curvature for physical values for the quark masses, namely
  $\langle r^4 \rangle = 0.73 \pm 0.09$~fm$^4$
  or equivalently $c_V=4.00\pm 0.50$ GeV$^{-4}$.
\end{abstract}

\end{frontmatter}

\section{Introduction and notation}
The expectation value of the vector current 
between two pion fields may be written as
$$
\langle \pi^\pm(p') \ar j^\mu \ar \pi^\pm(p) \ra = (p+p')^\mu
F_\pi^V(q^2 )\ .
$$
Since the only form factor that we discuss is the charged pion form
factor, we will denote it simply as $F(q^2)$, where $q=p-p'$. It is conventionally
normalized as $F(0)=1$.

An expansion around zero momentum transfer allows for a physical
interpretation of the form factor in terms of the pion's rest frame
charge density $\rho(r)$, given by
\be
 \label{Fexpansion}F(t)= 1+
\frac{1}{3!} \langle r^2\rangle_\rho t + \frac{1}{5!} \langle
r^4\rangle_\rho t^2+ {\mathcal O}(t^3) \ . \ee
 Here we used the notation of
Ref.~\cite{hofstaedter}.  Alternatively, in Ref.~\cite{Gasser:1990bv}
the curvature of the form factor was introduced via
\be\label{Fexpansion2}%
F(t)=1+\frac{1}{6}\la r^2 \ra t + c_V^\pi t^2 + {\mathcal O}(t^3) \ .  \ee%
Comparing with Eq. (\ref{Fexpansion}) we see that
$$
c_V^\pi = \frac{1}{5!} \langle r^4 \rangle \ .
$$

The first term in the form--factor expansion is the conventional
charge normalization $\int d^3r \rho(r)=\langle 1 \rangle_\rho = 1$,
and the derivative at the origin provides the (vector or charge)
pion  radius $\la r^2 \ra =6 (dF/dq^2)(0)$. At the next order, the
non-relativistic interpretation should be modified as effects of
boosting the pion wave function should begin to appear. In this
article we will ignore this subtlety, and simply use equivalently
the term pion's mean quartic radius or form factor curvature. We
will mainly focus on this quantity. Using both chiral perturbation
theory and dispersion relations we find a reliable value for the
curvature.

Assuming that the $\rho\pi\pi$ coupling is largely independent of
the quark masses for a given quark mass dependence of $m_\rho$, we
can also predict the quark mass dependence of the curvature.
The quark mass dependence of the $\rho$ properties
was studied in various recent lattice
simulations~\cite{Aoki:1999ff,Gockeler:2008kc} as well as 
using unitarized chiral perturbation theory~\cite{MadridJuelich}.

Abundant data on the pion form factor exists, that can be obtained
from the Durham reaction database. For the timelike form factor we
use contemporary sets from the CMD2, KLOE, and SND experiments
\cite{data:novosibirsk,Aloisio:2004bu,achasov}.
In addition there is
 higher energy data from Babar\cite{Solodov:2002xu} that
shows the $\rho(1700)$ and a shoulder that could correspond to the
$\rho(1400)$. However, in our study we employ only the $\rho(770)$.
We will therefore not extend our study beyond 1.2 GeV, where in
addition $K{\bar K}$ and other inelastic channels  start to
contribute significantly. It is this condition that prevents us from
studying the radius instead of the curvature, as will be explained
below.

In the case of the spacelike pion form factor the data is taken from
the CERN NA7 collaboration \cite{Amendolia:1986wj}. The more recent
data from JLAB~\cite{Volmer:2000ek} was taken at values of $Q^2$ too
large for our study. For a  recent review on the status of the
spacelike form factor see Ref.~\cite{Huber:2007zzb}.

The two-loop chiral perturbation theory ($\chi$PT) analysis of
\cite{Bijnens:2002hp} yielded a mean quadratic radius of
\be%
\langle r^2 \rangle = 0.452(13)\ {\rm
fm}^2 \ ,
\ee%
which is the currently accepted value~\cite{Yao:2006px}.  In order
to get a feeling on what to expect for the quartic radius, we start
with some simple classical examples. For this discussion we will
divide it by the mean square radius squared, the resulting ratio
$R\equiv\la r^4\ra/\la r^2\ra^2$ quantifies the radial spread of the
charge distribution. For a charge conducting sphere the spread is
minimal with $R = 1$ (all the charge is at the surface), and for a
uniformly charged dielectric sphere $ R = 25/21$. On the other hand,
the ratio is as large as 5/2 for a charge distribution with an
exponential dependence on the radius $e^{-mr}$. A vector-meson pole
form factor
$$
F(t) = \frac{1}{1-t/m_\rho^2}
$$
gives an even higher value of 10/3~\cite{Gasser:1990bv}.

Some results about the pion's quartic radius, including this work,
is collected in table \ref{tablequartic}.
\begin{table}[t]
\caption{Current theoretical estimates of the pion's quartic radius
\label{tablequartic}. The lattice results are our estimate based on
the form factor data from \cite{Boyle:2008yd}.}
\begin{center}
\begin{tabular}{|c|c|}
\hline
$\la r^4\ra/\la r^2\ra^2$ & Method \\
\hline
$3.3$ & VMD~\cite{Gasser:1990bv} \\
$2\pm 2.5$ & Lattice $m_\pi=0.33$~GeV~\cite{Boyle:2008yd}\\
$2 \pm 2.3$ & Lattice extrapolated to $m_\pi=0.139$~GeV~\cite{Boyle:2008yd}\\
$4 \pm 2$ & NNLO $\chi$PT~\cite{Bijnens:1998fm} \\
$3.1\pm0.4$ & Pad\'{e} approximants~\cite{Masjuan:2008fv} \\
$3.6 \pm 0.6$ & Eq. (\ref{curvatureq}) this work \\\hline
\end{tabular}
\end{center}
\end{table}
Furthermore, after having analyzed the pion em form factor data by
using analyticity, the curvature was constrained in the range
$[0.25~{\rm GeV}^{-4},7.57~{\rm GeV}^{-4}]$ in
Ref.~\cite{Caprini:1999ws} and $[2.3~{\rm GeV}^{-4},5.4~{\rm
GeV}^{-4}]$ in a very recent analysis~\cite{Ananthanarayan:2008mg}.
Of particular interest for us, and for a lattice determination, is
the quark mass dependence, or $m_\pi$ dependence, of the quartic
radius. A study of this within chiral perturbation theory would
require control of the N$^3$LO Lagrangian, since the quartic radius
is NNLO itself, and this seems out of today's reach.

We examine the problem with the simplifying assumption that
$g_{\rho\pi\pi}$ is $m_\pi$--independent while the $m_\pi$
dependence of $m_\rho$ is taken from other sources. To control the
model dependence, we employ the Omn\`es representation of the form
factor, sketched in Subsection \ref{Omnesintro} below. In the
absence of form factor zeroes, and neglecting inelastic channels,
this only requires knowledge of the elastic pion--pion scattering
phase shifts. We parametrize them, with a simple Breit--Wigner model
described in Subsection \ref{BWsubsection}. Since this model
contains $g_{\rho\pi\pi}$ and $m_\rho$ as the only parameters, the
above assumption can be employed in a straightforward way. In
addition we also use the predictions of unitarized chiral
perturbation theory for both rho mass and coupling. The resulting
quark mass dependence of the rho properties were investigated in
Ref.~\cite{MadridJuelich}. Using this alternative parametrization we
get almost identical results. The pion mass dependence of the
curvature turns out to be similar to that of the square radius.

\section{Omn\`es representation of the form factor}
\subsection{Basics}\label{Omnesintro}
The Omn\`es equation \cite{Omnes:1958hv} encodes the analyticity
properties of the pion form factor $F(s)$, that has an elastic
unitarity cut on the positive $s$-axis for $s\in(4m_\pi^2,\infty)$,
and is otherwise analytic. Further superimposed cuts due to
inelastic channels are neglected in its derivation, and the form
factor is assumed to have no zeroes (which, as we know today, is
phenomenologically correct). We have explored the possibility of
zeroes in the complex plane by analytically continuing the
experimental data with the help of the Cauchy--Riemann equations
\cite{GimenoSegovia:2008sx}. For a small band around the real axis,
they can be excluded. Some remarks on inelastic channels can be
found in \cite{Leutwyler:2002hm}.

The starting point is the well known relation ${\rm
Im}(F)=\tan\delta_{11} {\rm Re}(F)$, which relates the discontinuity
in the vector form factor to the elastic scattering phase shift in
the vector--isovector channel. From this relation the Watson theorem
follows straightforwardly. Since the  large-$q^2$ asymptotic
behavior of the form factor is known from QCD counting
rules~\cite{brodskyfarrar}, $F(q^2)\to c/q^2$, as a matter of
principle one may write an unsubtracted dispersion relation, which
reads for arbitrary $t$ \be \label{omnes} F(t)=\frac{1}{\pi}
\int_{4m_\pi^2}^\infty ds \tan \delta_{11}(s)
 \frac{{\rm Re}(F(s))}{s-t-i\epsilon}\  .
\ee%
We specified ``as a matter of principle'' since the QCD counting
rules apply only when elastic scattering is irrelevant by
the numerous inelastic channels open. However, in this work we only
want to use low energy input (up to 1.2 GeV) and we will therefore
use a subtracted dispersion relation below and cut the high energy
contributions with a cut--off.
 The
variation of the results with this cut--off provides a
systematic uncertainty in our work, which, as a consequence of
the subtraction, turns out to be moderate.

If there are no  bound state poles, as is the case of $\pi\pi$
scattering for physical quark masses, nor the form factor vanishes
anywhere in the complex plane, as we presume for $F(t)$, the
celebrated solution family of this equation provides a
representation of the form-factor in terms of the scattering phase,
known as the Omn\`es representation. The standard treatment proceeds
by deriving an integral equation for $\log F(t)/(2i)$ instead of
$F(t)$ itself, \ba \label{logomnes} \log \frac{F(t)}{2i} =
\frac{1}{2\pi i} \int_{4m_\pi^2}^\infty \frac{ds}{s-t} \left( \log
\frac{F(s+i\epsilon)}{2i} - \log \frac{F(s-i\epsilon)}{2i} \right)
 = \frac{1}{\pi} \int_{4m_\pi^2}^\infty \frac{ds}{s-t}
\delta_{11}(s) \ . \ea
Instead of this relation we may use a subtracted
version. This will allow us to effectively suppress
the high energy behaviour of the phase shifts. In particular we will use a twice subtracted version
which reads after exponentiation
%
\be\label{omnesrep2} %
F(t) =\exp\left(P_1 t+\frac{t^2}{\pi} \int_{4m_\pi^2}^\infty ds
\frac{\delta_{11}(s)}{s^2(s-t-i\epsilon)} \right) \ .
\ee%
Note, the normalization condition of the form factor
prohibits a constant term in the exponent.
The constant $P_1$ can be identified with the
square radius of the pion
$$
P_1 = \langle r^2\rangle /6 \ .
$$
This
representation of the form factor has been used in
the literature, see for example \cite{guerrero}.

\subsection{Phase--shift representation of the quartic radius}

Recalling the definition of the curvature of the pion form factor
(c.f. Eq.~(\ref{Fexpansion2})) we may read off an expression for
$c_V^\pi$ directly from Eq.~(\ref{omnesrep2}):
%
\be %
\label{curvatureq} c_V^\pi = \frac{\langle r^4 \rangle}{120} =
\frac{1}{72} \langle r^2 \rangle^2 + \frac{1}{\pi}
\int_{4m_\pi^2}^\infty ds  \frac{\delta_{11}(s)}{s^{3}}
\ee %
which is quite a beautiful formula, since it allows a third
independent extraction of the curvature $c_V^\pi$
besides
 NNLO $\chi$PT or a fit to spacelike data beyond
the linear fall in $t$ where uncertainties get large.
Instead we employ only
the elastic phase shift.
In addition, since the quantity
\be%
\label{curvaturefromphase} \tilde{c}_V^\pi\equiv c_V^\pi -
\frac{1}{72} \langle r^2 \rangle^2
\ee%
is described solely in terms of the $\pi\pi$ $p$--wave phase shifts,
its quark mass dependence is closely linked to that of the
$\rho$--meson properties. This relation is analogous to others
existing for the mean--square radius~\cite{Gasser:1990bv,Oller:2007xd}.

\subsection{A simple Breit--Wigner model} \label{BWsubsection}

To provide an estimate of the form factor based on the Omn\`es
representation, we employ a simple relativistic Breit--Wigner model
of the scattering amplitude, in which an $s$--channel resonance
dominates the scattering \be a_{11}(s) =
\frac{c}{s-m_\rho^2-im_\rho\Gamma_{\rm tot}(s)} \ee where
$\Gamma_{\rm tot}$ is the total width of the $\rho$ resonance with
\be \Gamma_{\rm tot} = \frac{g_{\rho\pi\pi}^2 p^3}{6\pi m_\rho^2} =
\frac{g_{\rho\pi\pi}^2 (\frac{s}{4}-m_\pi^2)^{3/2}}{6\pi m_\rho^2}\
. \ee We neglected terms of order $\Gamma_{\rm tot}^2$.
 Here $c$ is
an irrelevant constant that may be expressed in terms of $m_\rho$
and $\Gamma_{\rm tot}$. We may write
\be%
\delta_{11}(s) = \arctan \frac{{\rm Im} a_{11}(s)}{{\rm Re}
a_{11}(s)}  = \arctan \frac{m_\rho \Gamma_{\rm
tot}(s)}{m_\rho^2-s} \ .
\ee%
With this phase variation the integral representation converges
without additional subtractions (although the high--energy tail is
ad-hoc), however, even values as large as $s=7$ GeV$^2$ contribute
to the integral.

\begin{center}
\begin{figure}[t]
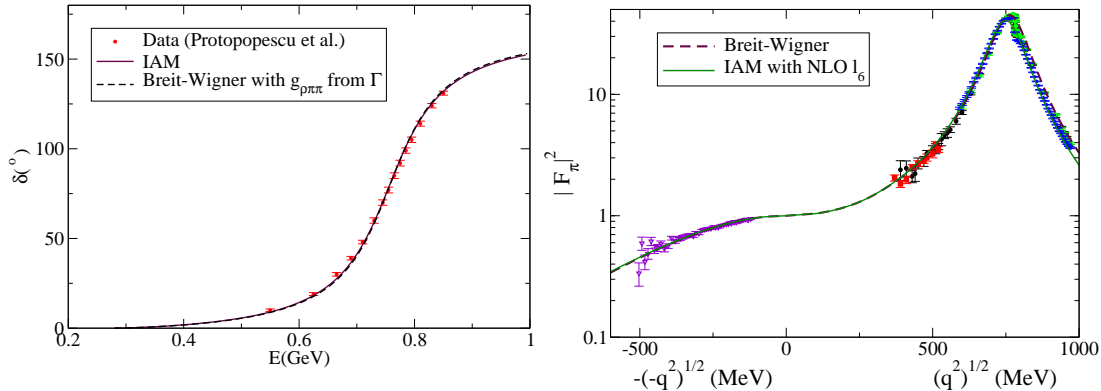

\parbox{7cm}{\includegraphics[width=7cm,angle=0]{Rhophase_notitle.eps}
\vspace{0.01cm}}
\parbox{7cm}{\includegraphics[width=7.4cm,angle=-0]{Formfactortimespace_notitle.eps}
\vspace{0.02cm}}
\caption{The scattering phase in the vector channel (left) for the
Breit--Wigner model (dashed line) and the Inverse Amplitude Method
(solid line). We also plot the square form factor modulus (right).
To be able to plot the spacelike and timelike data together, the
first is plotted against the unphysical variable $-\sqrt{-q^2}$ with
$q^2$ the (negative) spacelike momentum transfer. Here we use
$\langle r^2\rangle$ as input as described in the text.
 \label{fig:phase}}
\end{figure}
\end{center}

A good fit can be seen in Fig.~\ref{fig:phase} for the phase shift
and square form--factor modulus. To produce the figures we use
$m_\pi^{\rm phys}=139$~MeV, $m_\rho=775$~MeV, $\Gamma_\rho^{\rm
phys}=150$~MeV (to determine $g_{\rho\pi\pi}$).

Using the formula given above we may now extract
the curvature directly from the elastic pion
phase shifts, using the square radius as input.
We find
\be%
\label{eq:cvno}
c_V^\pi = 3.75\pm0.33 \ \mbox{GeV}^{-4} \ ,
\ee%
where the uncertainty contains both the uncertainty in $\langle
r^2\rangle$ and the systematic uncertainty introduced by evaluating
the integral only up to finite values (we allow a large range from
1~GeV to 16~GeV for the variation of the cut--off, although the
integral is basically converged for a cut--off of 2 GeV). It agrees to
that from vector meson dominance which is about
3.5~GeV$^{-4}$~\cite{Gasser:1990bv} and it is consistent with the
constraint $[2.3~{\rm GeV}^{-4},5.4~{\rm GeV}^{-4}]$ from analyzing
the form factor data using analyticity~\cite{Ananthanarayan:2008mg}.
The advantage of our analysis is that it allows in addition for a
controlled estimate of the uncertainty. Equivalently, the result in
term of quartic radius is
\be%
\la r^4 \ra = 0.68\pm0.06~{\rm fm}^4 \ .
\ee%

As mentioned above we will investigate the quark mass dependence of
the pion form factor based on the assumption that $g_{\rho \pi\pi}$
is independent of the quark mass with the $m_\pi$ dependence of
$m_\rho$ taken from other sources. Since both parameters are
explicit in the parametrization given above, we may study the
resulting quark mass dependence of $c_V^\pi$, once that of $\langle
r^2\rangle$ is fixed.

\section{Chiral perturbation theory}
\subsection{General considerations\label{subsec:defNNLO}}

In order to determine the quark mass dependence of the square
radius, which is the input needed for the formalism described above,
we will use the results of $\chi$PT. Clearly, the curvature
$c_V^\pi$ as well as its quark mass dependence, could also be
determined in $\chi$PT directly. Depending on the fit and
systematics chosen in Ref.~\cite{Bijnens:1998fm}, which is
two--flavor ${\mathcal O}(p^6)$ $\chi$PT calculation, its value could vary
between $2-6$ GeV$^{-4}$, although the authors quote a value around
4 GeV$^{-4}$, in agreement with a previous estimate
\cite{Gasser:1990bv} (By fitting to form factor data, they obtain
$3.85$~GeV$^{-4}$). A ${\mathcal O}(p^6)$ fit in three--flavor $\chi$PT leads
to a range $4.49\pm0.28$~GeV$^{-4}$~\cite{Bijnens:2002hp}. Adopting
$c_V^\pi=4\pm 2$~GeV$^{-4}$ as the NNLO $\chi$PT result, we obtain
$\langle r^4 \rangle/ \langle r^2 \ra^2= 4\pm2$. This value is
copied into Table~\ref{tablequartic}.

\subsection{Matching the Omn\`es representation}
We start by giving the chiral expansion of the vector form factor
\cite{gasserleutwyler} valid to NLO in $\chi$PT,
\be \label{ffinchipt}%
F(t) = 1 + \frac{1}{6f_\pi^2}(t-4m_\pi^2)\bar{J}(t)
+\frac{t}{96\pi^2f_\pi^2}(\bar{l}_6-\frac{1}{3}) \ .
\ee%
(To this order we are free to change  $M,\ F$ to
the physical $m_\pi, f_\pi$, since the difference is of  NNLO). In
this expression,
\ba%
\bar{J}(t) = \frac{1}{16\pi^2} \left[ \sigma \log \left(
\frac{\sigma-1}{\sigma+1}\right)+2 \right]
\ea%
with $\sigma = \sqrt{1-4m_\pi^2/t}$. A common strategy is to fix the
$\bar{l}_6$ constant from the square charge radius
\cite{gasserleutwyler}
\be \label{radio}%
\la r^2 \ra = \frac{1}{16\pi^2f_\pi^2} (\bar{l}_6-1)
\ee%
which is correct up to ${\mathcal O}(m_\pi^2)$ in $\chi$PT. Higher orders in
the chiral expansion cannot bring in powers of $t$ since, by
definition, the charge squared radius is proportional to the
coefficient of the term linear in $t$ in the form factor. However,
they can bring additional constants  to the right hand side (each of
a natural order suppressed by additional factors of $1/(4\pi
f_\pi)^2$), and, more important for our purposes, a polynomial of
$m_\pi^2$. To make sure we are not eschewing a critical $m_\pi$
dependence, we will compare the right-hand-side of eq.(\ref{radio})
with the NNLO correction in chiral perturbation theory
\cite{Bijnens:1998fm}. The NLO result eq. (\ref{radio}), that
depends only logarithmically on the pion mass (see
eq.~(\ref{eq:lbarmpi}) below), is then extended to
\ba %
\langle r^2 \rangle=
\frac{1}{16\pi^2f_\pi^2} \left[ \left( 1 +
\frac{m_\pi^2}{8\pi^2f_\pi^2}\bar{l}_4\right) (\tilde{l}_6-1) +
\frac{m_\pi^2}{16\pi^2f_\pi^2}\left(
16\pi^2\frac{13}{192}-\frac{181}{48}\right) \right] \ea %
with
\be%
\label{tildeseis} \tilde{l}_6=\bar{l}_6 + 6 \frac{m_\pi^2}{f_\pi^2}
\left[ 16\pi^2 r^r_{V1}(\mu^2)+ \frac{1}{48\pi^2}\log \left(
\frac{m_\pi^2}{\mu^2}   \right) \left(
\frac{19}{12}-\bar{l}_1+\bar{l}_2\right) \right]
\ee %
where
$r^r_{V1}$ is a counterterm to be determined empirically, and we
will use the simple VMD estimate from the same work, at the $\rho$
scale,
$$
r^r_{V1}(m_\rho^2) \simeq -0.25 \times 10^{-3} \ .
$$
With this estimate, those authors find
$$
\tilde{l}_6 = \bar{l}_6 -1.44
$$
(the scale--dependence of this number cancels in
Eq.~(\ref{tildeseis}). The estimate is taken with constants
corresponding to set I that we copy in Table
\ref{tabla:constantes}).

Here we have to recall the pion mass dependence of the $\bar{l}$'s.
The $l_i$, as coefficients of the expansion in powers of $m_\pi^2$
of the Lagrangian density, are by definition pion--mass independent,
and so are their renormalized counterterms $l_i^r$. However, the
barred quantities are related to them by absorbing a chiral
logarithm
\be%
l^r_i = \frac{\gamma_i}{32\pi^2} \left[ \bar{l}_i+\log\left(
\frac{m_\pi^2}{\mu^2}\right) \right]
\ee%
that makes the $\bar{l}$'s scale--independent, but in exchange,
pion--mass dependent. This dependence needs to be kept track of in
the calculation. It becomes crucial in the chiral limit when the
pion radius diverges due to the virtual pion cloud becoming
long--ranged as the pion mass vanishes. This effect appears through
$\bar{l}_6$.

Therefore, we denote by $\bar{l}_i^{\rm phys}$ the value that the
low energy constants take by fitting to physical--world data. From
here on, when varying the quark (or pion) mass, one needs to change
the constant according to
\be%
\label{eq:lbarmpi} \bar{l}_i = \bar{l}_i^{\rm phys} - \log \left(
\frac{m_\pi^2}{(m_\pi^{\rm phys})^2} \right)
\ee%
With this we have all the input ready to use Eq.~(\ref{curvatureq})
also to establish the quark mass dependence of the curvature using
the Breit--Wigner representation of the phase shifts and the given
assumptions on the quark mass dependence of both the $\rho$ mass and
coupling.

However, before we proceed we introduce a method that allows one to
estimate the quark mass dependence of the $\rho$ properties directly
from the $\chi$PT amplitudes evaluated up to a given order, namely
the unitarized chiral perturbation theory or the inverse amplitude
method (IAM). The representation we are going to use is consistent
with NLO chiral perturbation theory at low momentum, and satisfies
exact elastic unitarity, fitting the pion scattering data up to
1.2~GeV well. The formalism will be introduced in the next
subsection.

\begin{figure}[t]
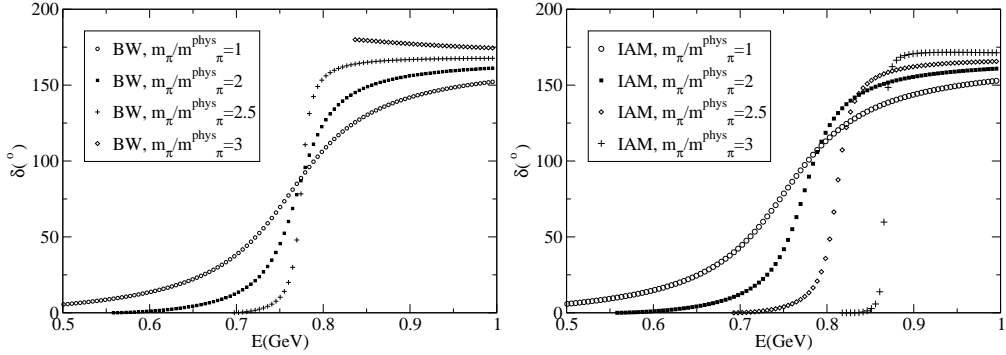

\begin{center}
\hglue-0.8cm\includegraphics[width=0.4\textwidth]{phaseofmBW_notitle.eps}
\hglue-0.0cm\includegraphics[width=0.4\textwidth]{phaseofmIAM_notitle.eps}
\end{center}
\vglue-0.2cm \caption{Variation of the elastic $\pi\pi$ phase
$\delta_{11}$ with the pion mass. Left: Breit--Wigner model. Note
that for $m_\pi= 3m_\pi^{\rm phys}$ the $\rho$ (held at constant
mass) has already crossed below the $\pi\pi$ threshold and is a
bound state. Right: Inverse Amplitude Method. The resonance stays
above the $\pi\pi$ threshold, its mass having a slight dependence on
$m_\pi$, until rather high pion masses. \label{fig:phaseBW}}
\end{figure}

\subsection{P-wave $\pi\pi$ scattering}

To derive the expression for the IAM one starts with the on--shell
$\pi\pi$ scattering amplitude in NLO $\chi$PT that, for  $I=1$, is
\be A_1(s,t,u)= A(t,s,u)-A(u,t,s) \ee with
\begin{eqnarray} \label{pipiamplitude} \nonumber
A(s,t,u)&=& \frac{s-m_\pi^2}{F^2} + \frac{1}{6F^4}\left[3\bar{J}(s)
\left(s^2-m_\pi^4 \right)+
\bar{J}(t)\left(t(t-u)-2m_\pi^2t+4m_\pi^2u-2m_\pi^4\right)\right.
\\ \nonumber
&+&\left.
\bar{J}(u)\left(u(u-t)-2m_\pi^2u+4m_\pi^2t-2m_\pi^4\right)
\right]  + \frac{1}{96\pi^2f_\pi^4} \left[ 2\left(\bar{l}_1-\frac{4}{3}\right)
\left(s-2m_\pi^2\right)^2
 \right. \\  &+&\left.
\left(\bar{l}_2-\frac{5}{6}\right)\left(s^2+(t-u)^2\right) -3
m_\pi^4\bar{l}_3 -12m_\pi^2s +15m_\pi^4 \right] .
\end{eqnarray}
The first term can be identified as the leading order, low--energy
theorem \cite{weinberg}, but we express it in terms  of the physical
$m_\pi$, instead of its leading order value $M$ used in the original expression
\cite{gasserleutwyler}. At the meanwhile, we keep the $m_\pi$
independent pion decay constant $F$. The quantities $F
$ and $M$ are related to the physical ones via
$$
F=f_\pi\left(1-\frac{m_\pi^2}{16\pi^2f_\pi^2} \bar{l}_4 \right), \ \
\ M^2=m_\pi^2\left(1+\frac{m_\pi^2}{32\pi^2f_\pi^2} \bar{l}_3
\right).
$$
The latter expression introduces $\bar{l}_3$ into the last line of
Eq.~(\ref{pipiamplitude}).

The projection to the spatial $p$--wave has the usual factor of
$1/2$ to avoid double--counting quantum states by counting  all
angular configurations with exchanged identical particles
\be%
a_{11}(s) = \frac{1}{32\pi} \frac{1}{2}  \int_{-1}^1 d\cos \theta
(\cos \theta) A_1(s,t(s,\cos \theta),u(s,\cos \theta)) \ .
\ee%
One can organize the chiral expansion as
\be%
a_{11}(s) = a_{11}^{\rm LO}(s) + a_{11}^{\rm NLO}(s) +\ \dots
\ee%
but the series truncated at any order only satisfies elastic unitarity
perturbatively.
This is solved, with the  first two expansion terms, by
the Inverse Amplitude Method \cite{truong} that reads (suppressing
the spin and isospin subindices)
\be%
a^{\rm IAM}(s) = \frac{a^2_{\rm LO}(s)}{a^{\rm LO}(s)-a^{\rm
NLO}(s)} \ .
\ee%
A Taylor expansion of this amplitude returns NLO $\chi$PT as usual
for a Pad\'e approximant. However elastic unitarity is now exact,
and the possibility   of  a zero of the denominator allows for
resonances to appear.

The associated phase shift
$$\delta_{11}^{\rm IAM}(s)= \arctan \left( \frac{{\rm Im} a_{11}^{\rm IAM}(s)}{ {\rm
Re} a_{11}^{\rm IAM}(s)} \right)
$$
may be directly employed for the time--like form factor through the
Omn\`es representation. A similar procedure was taken to calculate
the scalar and vector form factors of the
pion~\cite{Truong:1988zp,Guerrero:1998ei}.

The pion mass dependence of the $\rho$ meson properties were studied
in Ref.~\cite{MadridJuelich} and it was found that $g_{\rho \pi\pi}$
depends only very mildly on the quark mass. In the next section we
will investigate the consequences of this finding on the pion vector
form factor.

The low energy constants necessary to complete the calculation are
fit to the phase shifts data and given in
Table~\ref{tabla:constantes}, where they are compared to well--known
determinations. Note that with the phase shift data one can only
determine the difference ${\bar l}_2-{\bar l}_1$ which is about 6 as
a result of the fitting~\cite{Dobado:1996ps}. Using
Eq.~(\ref{curvatureq}), the curvature can then be obtained as
\be%
c_V^\pi = 4.00\pm0.50~{\rm GeV}^{-4},
\ee%
or equivalently,
\be%
\langle r^4\rangle = 0.73\pm0.09~{\rm fm}^{4}.
\ee%
The quantity depending solely on the phase shift is
\be%
\tilde{c}_V^\pi = 2.13\pm0.42~{\rm GeV}^{-4}.
\ee%
These values are to be considered as our results at the physical
pion mass.
\begin{table}[t]
\caption{ Values of the low energy constants of the NLO SU(2) Chiral
Lagrangian. We employ the last row in the calculation. For
comparison we give several well--known sets. The error refers to the
last significant figure. These are determinations based on data
alone. Several phenomenological and theoretical predictions based on
semi-analytical approaches (large $N_c$, Dyson--Schwinger, resonance
saturation, etc.) can be found in the literature \cite{various}.
\label{tabla:constantes} }
\begin{center}
\begin{tabular}{|c|c|c|c|c|c|}
\hline%
LEC                             &$\bar{l}_1$&$\bar{l}_2$
&$\bar{l}_3$ &$\bar{l}_4$ & $\bar{l}_6$
\\ \hline
Gasser-Leutwyler~\cite{gasserleutwyler}                &$-2 \pm 4$   &  $6\pm 1.3$  & $2.9\pm 2.4$ & $4.3\pm 0.9$  &
$16\pm 1$ \\
Dobado {\it et al.}~\cite{dgnmp} &$-0.6\pm 0.9$ & $6.3 \pm 0.5$   & $2.9\pm 2.4$ & $4.3\pm 0.9$  &
$16\pm 1$ \\
Bijnens {\it et al.} set I~\cite{Bijnens:1998fm}     & $-1.7$   & $6.1$      & $2.4$    & $4.4\pm 0.3$
&
$16\pm 1$  \\
Bijnens {\it et al.} set II~\cite{Bijnens:1998fm}        & $-1.5$   &  $4.5$     &  $2.9$   &
$4.3$ &
\\
This work                       & $0.1\pm 1.5$   & $6\pm 1.3$ & $2.9$(fix)
& $4.3\pm 0.9$ & $16.6\pm 0.4$ \\\hline
\end{tabular}
\end{center}
\end{table}

\section{Pion-mass dependence needed for lattice extrapolation}
\subsection{Mass dependence of the phase shift} \label{subsec:mpidep}
In this section we study the pion mass dependence of the pion vector
form factor based on both the Breit--Wigner model as well as the
amplitudes from the IAM. In the left panel of Fig.~\ref{fig:phaseBW}
we plot the variation in the isospin--1 $p$--wave elastic $\pi\pi$
phase shift $\delta_{11}$ with the pion mass in the Breit--Wigner
model, where the physical pion mass is denoted by $m_\pi^{\rm phys}$. For
small increases in the pion mass, with $m_\rho$ being held fixed for
illustration, we see how the resonance becomes narrower as the pion
threshold approaches. Finally, for $2m_\pi>m_\rho$, the $\rho$
becomes bound and the phase shift starts at 180 degrees in agreement
with Levinson's theorem with one bound state.

Next we consider the IAM. Here what is held constant is the
renormalized constants in the chiral Lagrangian ($l_i^r(\mu)$)
since, as discussed, they are by definition independent of the pion
mass. The scale--independent $\bar{l}'s$ run logarithmically with
the pion mass. This dependence and the explicit pion masses in the
chiral series bring about a small $m_\pi$--dependence of the $\rho$
mass that puts it just above threshold for $m_\pi=3m_\pi^{\rm
phys}$. We plot in the right panel of Fig.~\ref{fig:phaseBW} the
resulting phase as a function of the $\pi\pi$ invariant mass for
different values of the pion mass.

\begin{figure}[t]
\begin{center}
\includegraphics[width=0.58\textwidth]{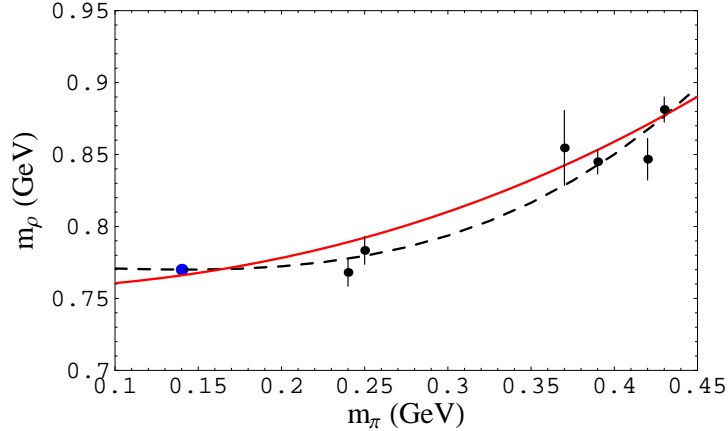}
\caption{Dependence of the $\rho$ mass on the pion mass. Here the $\rho$ mass is
defined as the value of $\sqrt{s}$, where
the $\pi\pi$ $p$--wave phase shifts cross
90 degrees. Shown are the result from the IAM
(solid line) and a fit using Eq.~(\ref{brunsfit}) (dashed line) to the lattice
data, shown as solid dots~\cite{Gockeler:2008kc}. The lowest point is
the physical $\rho$ mass. \label{fig:mrho}}
\end{center}
\end{figure}

The prediction of the pion mass dependence of the $\rho$ mass resulting from the IAM
is shown as the solid curve in Fig.~\ref{fig:mrho}. The
parameters used are ${\bar l}_1=-0.08,{\bar l}_2=5.78,{\bar
l}_3=2.9$ and ${\bar l}_4=4.3$.
In this figure we also show the results of
a recent lattice study~\cite{Gockeler:2008kc}.
For our comparison we choose this one, for it
is the simulation where the lowest pion masses
are used.
To allow for a comparison with recent lattice data, here the $\rho$--mass is defined as the value 
of $\sqrt{s}$, where the $\pi\pi$ phase shift
is 90$^\circ$. The resulting numbers 
differ somewhat from those corresponding to
the real part of the pole position in
the second Riemann sheet --- the latter definition
of the mass was used in Ref.~\cite{MadridJuelich}. For comparison, we also show the very
recent lattice data~\cite{Gockeler:2008kc} as the 
filled circles
with error bars. The agreement of the IAM with the lattice data is
rather satisfying. For later use, the lattice data are fitted with
an expression derived from an extended version of $\chi$PT~\cite{Bruns:2004tj}
\be%
m_\rho = m_\rho^0 + c_1m_\pi^2 + c_2m_\pi^3 +
c_3m_\pi^4\log\left({m_\pi^2\over m_\rho^2}\right) \ .
\label{brunsfit}
\ee%
The parameter $m_\rho^0$ is not included in the fit. It is fixed by the condition
that $m_\rho=0.77$~GeV at the physical pion mass. 
We find for the $\rho$ mass in the chiral limit $m_\rho^0=0.77\pm 0.1$ 
GeV.
The
resulting parameters are
\be%
c_1 = -0.53\pm0.44~{\rm GeV}^{-1}, \quad c_2 = 2\pm1~{\rm
GeV}^{-2}, \quad c_3 = -1\pm 3~{\rm GeV}^{-3}.
\ee
Using the central values, we get the dashed curve as shown in
Fig.~\ref{fig:mrho}.
Note, the uncertainties in the parameters show some correlation,
however, since we are here mainly interested in a parametrization
of the lattice data, we may ignore this observation.
 Since the pion mass grows faster with the quark
mass, eventually the $\rho$ becomes bound (just as the $J/\psi$ is
under the $D\bar{D}$ threshold), but this happens for yet larger
pion masses.

\subsection{Extrapolation in NLO and NNLO chiral perturbation theory}
Space--like form factors are in principle accessible on a lattice.
Since these studies usually employ heavier-than-real quarks, the
pion mass obtained is also higher than the physical pion mass, and
an extrapolation is necessary. Another extrapolation to low momentum
(due to the finite volume enclosed by the lattice) is necessary if
the mean square and quartic radii are to be extracted.  The mean
square radius has indeed been studied before
\cite{Boyle:2008yd,Bunton:2006va} and extrapolation to physical pion
masses taken from chiral perturbation theory. It would be
interesting to have lattice data at several quark masses to test it.

Momentum extrapolations to $q^2=0$ are, in view of the mean quartic
radius, non-linear. In the extraction of the mean square radius, the
authors of \cite{Bunton:2006va} quote a 10\%  systematic error in
the lattice extraction due to $m_\pi^2/(1~{\rm GeV}^2)$ $\chi$PT
errors, and 20\% due to $q^2_{\rm min}/(1~{\rm GeV}^2)$ momentum
extrapolation errors.

The momentum extrapolation however seems to be avoidable with twisted
boundary conditions for the fermion fields \cite{Boyle:2008yd}, and
indeed
those authors find
$$
\la r^2\ra_{330\ \rm MeV} = 0.35(3) \ {\rm fm}^2,\ \ \ \la
r^2\ra_{139\ \rm MeV} = 0.42(3) \ {\rm fm}^2\ .
$$
where the value at the physical pion mass is obtained with the help
of the NLO $SU(2)$ chiral Lagrangian.

\subsection{Chiral extrapolation assisted by the Omn\`es representation}

We have achieved a representation of the form factor based on the
Omn\`es representation, matched to low energy $\chi$PT. Since we
have relatively good theoretical control of the entire construction,
we can now extrapolate to unphysical quark (pion) masses.

The parametrization in Eq. (\ref{curvatureq}) requires two pieces of
input: the pion scattering $p$--wave phase shift and the mean square
radius. For the former we may either use the Breit--Wigner model --
together with additional assumptions on the $\rho$ properties -- or
the IAM, where the quark mass dependence is predicted from NLO
$\chi$PT -- higher order pion mass dependencies as they arise from
NNLO $\chi$PT are not yet studied in this framework.

\begin{figure}[t]
\begin{center}
\includegraphics[width=7cm]{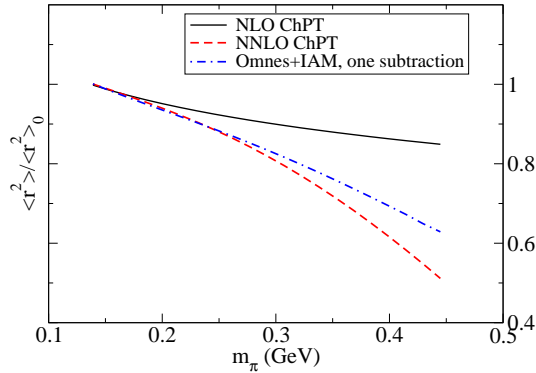}
\caption{Pion mass dependence of the mean square radius. We show
  results based on NLO and NNLO Chiral Perturbation theory. The
  $r_{V1}^r$ parameter is fixed at its VMD value (its pion-mass
  dependence contributing at NNNLO is neglected). Likewise we plot the
  mass dependence resulting from a once-subtracted Omn\`es
  representation. The data is normalized to the radius for physical
  pion mass. \label{fig:r2mass}}
\end{center}
\end{figure}

The square radius has a NLO pion mass dependence caused by the
chiral logarithm in $\bar{l}_6$. This is a major effect for pion
masses smaller than physical, towards the chiral limit, but for pion
masses higher than physical (say the 330~MeV where the lattice data
is taken), the $m_\pi^2$ term from the NNLO Lagrangian density might
come to dominate, so we employ this too. Finally, we have an
order-of-magnitude countercheck at our disposal. By employing a
once--subtracted instead of a twice--subtracted Omn\`es
representation, we obtain a closed form for the mean square radius
in terms of the phase shift
\be \label{radiusfromshift} %
\la r^2 \ra = \frac{6}{\pi} \int_{4m_\pi^2}^\infty ds
\frac{\delta_{11}(s)} {s^2} \ .
\ee %
All three methods are plotted in Fig.~\ref{fig:r2mass}.

\begin{figure}[t]
\hglue-8mm\includegraphics[width=0.6\textwidth]{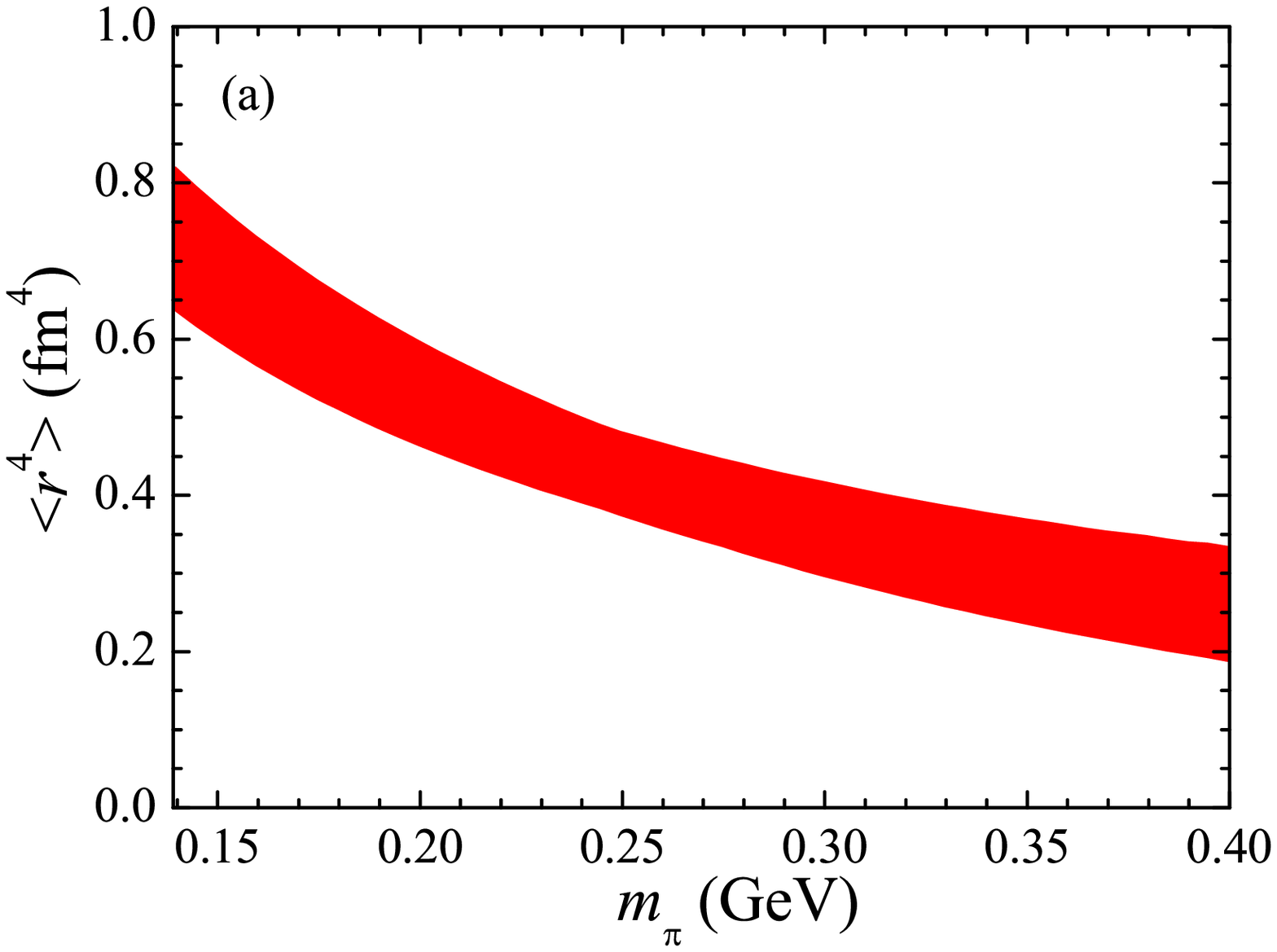}
\hglue-1.5cm\includegraphics[width=0.6\textwidth]{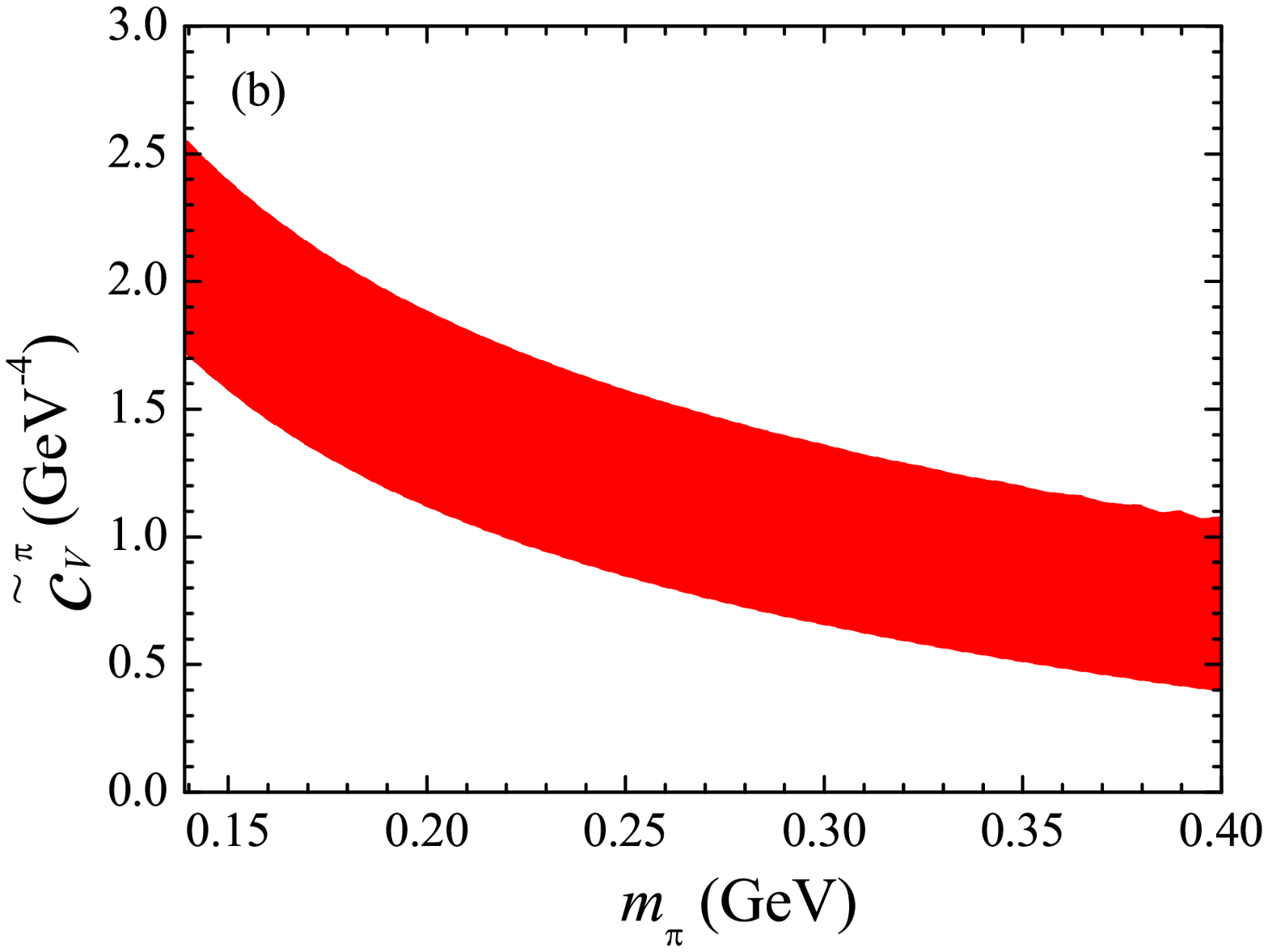}
\vglue-5mm
\hglue-8mm\includegraphics[width=0.6\textwidth]{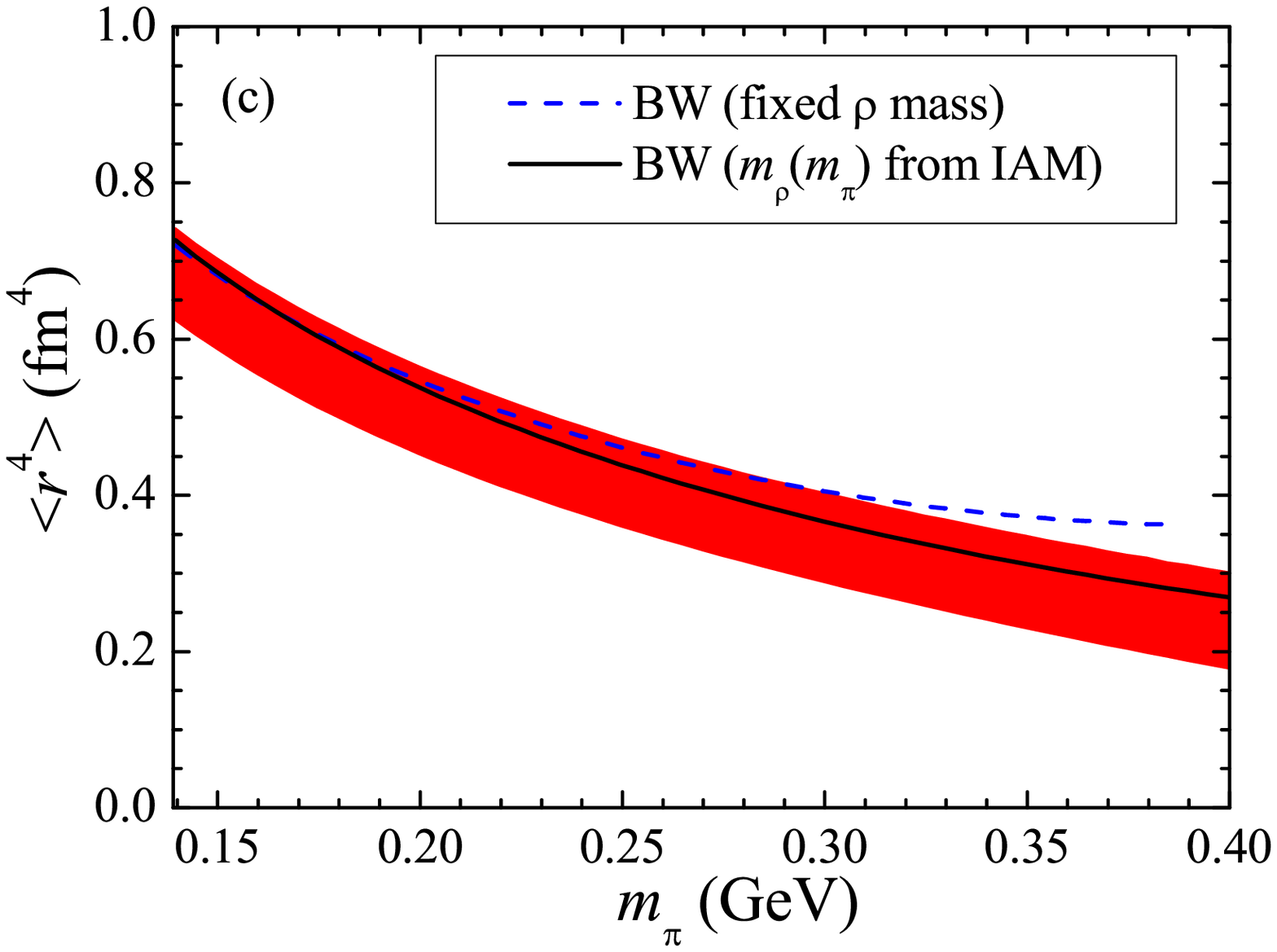}
\hglue-1.5cm\includegraphics[width=0.6\textwidth]{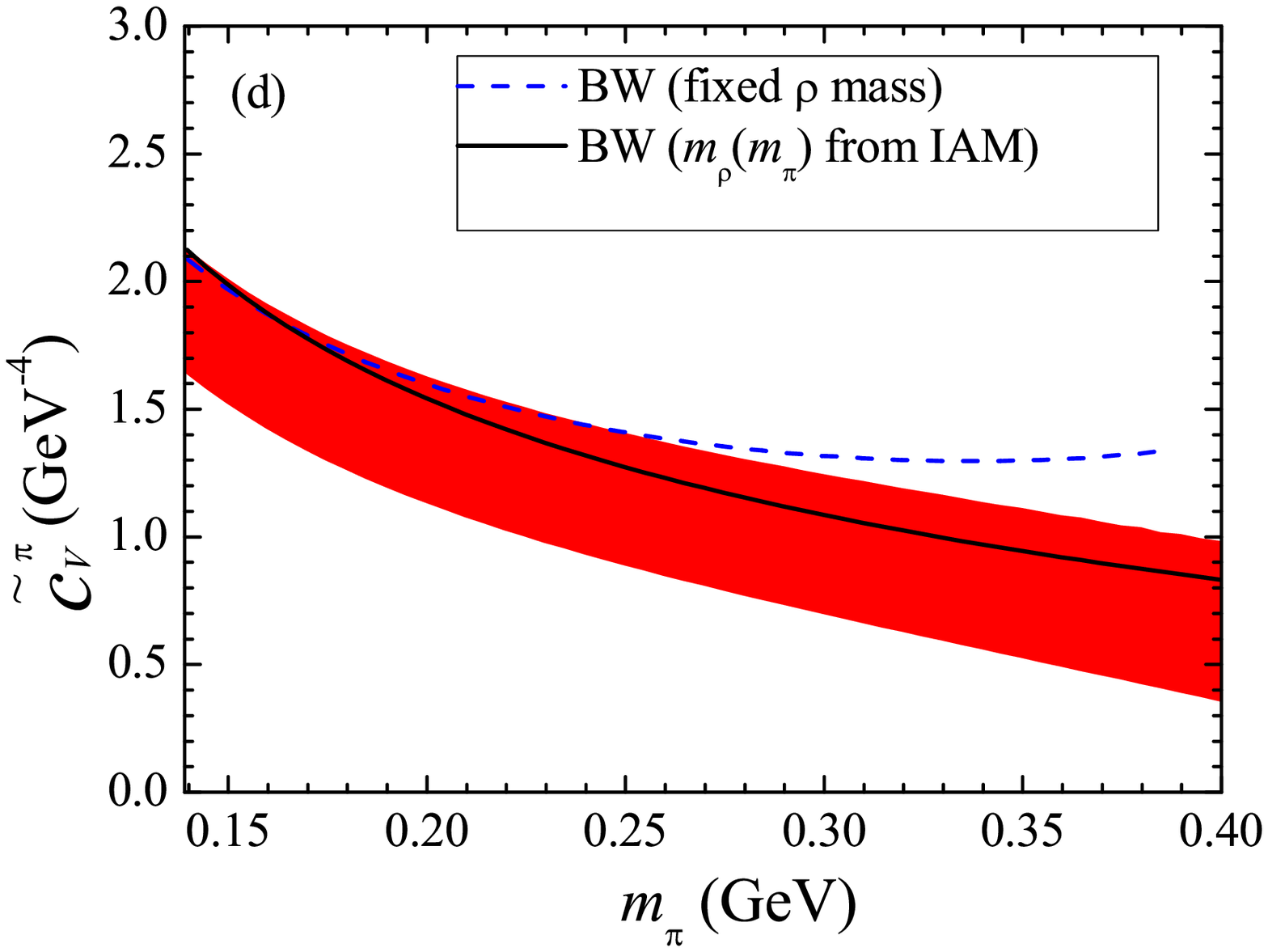}
\caption{Dependence on the pion mass of the mean quartic radius of the
  pion (left panel) and $\tilde{C}_V^\pi$ (right panel). We show
  results based on the Breit--Wigner model and the Inverse Amplitude
  Method. The bands correspond to the uncertainties from the
  parameters used (the ${\bar l}_i$'s for the IAM and the $c_i$ (c.f.
  Eq.~(\ref{brunsfit})) for the Breit-Wigner) as well as from the
  variation of the cut--off.\label{fig:r4mass}}
\end{figure}

The results of the pion mass dependence of the quartic radius using
the Breit--Wigner model and the IAM are plotted in
Fig.~\ref{fig:r4mass}(a) and (c), respectivley, with the $m_\pi$
dependence of the square radius coming from that of ${\bar l}_6$ as
dictated by Eq.~(\ref{eq:lbarmpi}). In the IAM, the pion mass
dependence of $m_\rho$ is included intrinsically, while in the
Breit--Wigner method, it is input from fitting to the recent lattice
data~\cite{Gockeler:2008kc} as described at the end of
Subsection~\ref{subsec:mpidep}. The bands include the uncertainty
from varying the parameters within one sigma and that from varying
the integration cut--off from 1 to 16~GeV. The uncertainty from the
cut--off is the dominant one. For comparison, we also plot the
result of the Breit--Wigner model with fixed $\rho$ mass as the
dashed curve in Fig.~\ref{fig:r4mass}(c). The dependence is smooth
up to the point when the rho becomes stable.
Here the curve ends.
Imposing the $m_{\pi}$--dependent $\rho$ mass as that given by the
IAM, the result for the quartic radius in the Breit--Wigner model is
shown as the solid curve in Fig.~\ref{fig:r4mass}(c).

We can dispose altogether from the explicit pion mass dependence in
$\la r^2\ra$ by studying the quantity
$$
\tilde{c}_V^\pi =  c_V^\pi - \frac{1}{72} \la r^2\ra^2 .
$$
This constant $\tilde{c}_V^\pi$ can of course be also studied on a
lattice by itself, although its physical interpretation is not
transparent. But its mass dependence comes from the phase shift
alone (c.f. Eq.~(\ref{curvaturefromphase})), and is not compounded
with the mass dependence of the square radius. It is therefore this
quantity that allows most directly access to the pion mass
dependence of the $\rho$ properties. Our results for this quantity
are shown in Fig.~\ref{fig:r4mass}(b) and (d) using the IAM and the
Breit--Wigner model, respectively.

\section{Summary}

Using the Omn\`es representation for the pion vector form factor, in
this paper we improved the existing value for the corresponding
curvature using as input only the well known $\pi\pi$ phase shifts
in the $p$--wave as well as the pion radius. 
We find $\langle r^4 \rangle = 0.73 \pm 0.09$~fm$^4$ 
or equivalently $c_V=4.00\pm 0.50$ GeV$^{-4}$ which are consistent
with the results from NNLO
$\chi$PT~\cite{Bijnens:1998fm,Bijnens:2002hp} and recent analysis
using analyticity~\cite{Ananthanarayan:2008mg}.

  In addition we studied the pion mass dependence of the curvature. A
  modification of the curvature, called $\tilde{c}_V^\pi$ in the paper,
  can be represented solely by the $\pi\pi$ $p$--wave phase shift. We
  argued that this quantity allows for a clean and model--independent
  alternative access to the pion mass dependence of the $\rho$ properties
  and would therefore provide a consistency check of the methods to
  extract physical parameters from lattice simulations. A lattice QCD
  study of the pion curvature would therefore be of high theoretical
  interest. We also argued that the pion square radius is not well suited for
  this kind of investigation, since additional, not so well
  controlled, theoretical input would be needed in the analysis.
  Quantities similar to $\tilde{c}_V^\pi$ exist also for other
  form factors, and a study of them from both theoretical and lattice sides
  would be interesting.



\vspace{0.5cm}

{\emph{ We would like to thank Stephan D\"urr
and Jose Pelaez for useful discussions.
This work is supported in part by grants FPA 2004-02602, 2005-02327,
FPA2007-29115-E (Spain), and by the Helmholtz Association through
funds provided to the virtual institute ``Spin and strong QCD''
(VH-VI-231). FJLE thanks the members of the IKP (Theorie) at
Forschungszentrum J\"ulich for their hospitality during the
preparation of this work and the Fundacion Flores Valles for
economical support. }}



\begin{thebibliography}{50}

\bibitem{hofstaedter}
 M.~E.~Rose,
  Phys.\ Rev.\  {\bf 73} (1948) 279.

\bibitem{Gasser:1990bv}
  J.~Gasser and U.-G.~Mei\ss ner,
  Nucl.\ Phys.\  B {\bf 357} (1991) 90.

\bibitem{Aoki:1999ff}
  S.~Aoki {\it et al.}  [CP-PACS Collaboration],
  Phys.\ Rev.\  D {\bf 60} (1999) 114508
  [arXiv:hep-lat/9902018];
    S.~D\"urr {\it et al.},
  Science {\bf 322} (2008) 1224;
   C.~Gattringer, C.~Hagen, C.~B.~Lang, M.~Limmer, D.~Mohler and A.~Schafer,
  arXiv:0812.1681 [hep-lat];
 P.~Dimopoulos, C.~McNeile, C.~Michael, S.~Simula and C.~Urbach  [ETM
                  Collaboration],
  arXiv:0810.1220 [hep-lat].
 
\bibitem{Gockeler:2008kc}
  M.~Gockeler, R.~Horsley, Y.~Nakamura, D.~Pleiter, P.~E.~L.~Rakow, G.~Schierholz and J.~Zanotti,
  arXiv:0810.5337 [hep-lat].

\bibitem{MadridJuelich}
  C.~Hanhart, J.~R.~Pelaez and G.~Rios,
  Phys.\ Rev.\ Lett.\  {\bf 100} (2008) 152001
  [arXiv:0801.2871 [hep-ph]].


\bibitem{data:novosibirsk}
    R.~R.~Akhmetshin {\it et al.},
  JETP Lett.\  {\bf 84} (2006) 413
  [Pisma Zh.\ Eksp.\ Teor.\ Fiz.\  {\bf 84} (2006) 491]
  [arXiv:hep-ex/0610016];
  R.~R.~Akhmetshin {\it et al.}  [CMD-2 Collaboration],
  Phys.\ Lett.\  B {\bf 648} (2007) 28
  [arXiv:hep-ex/0610021];

\bibitem{Aloisio:2004bu}
  A.~Aloisio {\it et al.}  [KLOE Collaboration],
  Phys.\ Lett.\  B {\bf 606} (2005) 12
  [arXiv:hep-ex/0407048].

\bibitem{achasov}
  M.~N.~Achasov {\it et al.},
  J.\ Exp.\ Theor.\ Phys.\  {\bf 101} (2005) 1053
  [Zh.\ Eksp.\ Teor.\ Fiz.\  {\bf 101} (2005) 1201]
  [arXiv:hep-ex/0506076].


\bibitem{Solodov:2002xu}
  E.~P.~Solodov  [BABAR collaboration],
in {\it Proc. of the $e^+ e^-$ Physics at Intermediate Energies
Conference } ed. Diego Bettoni,
{\it In the Proceedings of e+ e- Physics at Intermediate Energies, SLAC,
Stanford, California, 30 Apr - 2 May 2001, pp T03}
  [arXiv:hep-ex/0107027].


\bibitem{Amendolia:1986wj}
  S.~R.~Amendolia {\it et al.}  [NA7 Collaboration],
  Nucl.\ Phys.\  B {\bf 277} (1986) 168.

\bibitem{Volmer:2000ek}
  J.~Volmer {\it et al.}  [The Jefferson Lab F(pi) Collaboration],
  Phys.\ Rev.\ Lett.\  {\bf 86} (2001) 1713
  [arXiv:nucl-ex/0010009].

\bibitem{Huber:2007zzb}
  G.~M.~Huber,
  J.\ Phys.\ Conf.\ Ser.\  {\bf 69} (2007) 012015.


\bibitem{Bijnens:2002hp}
  J.~Bijnens and P.~Talavera,
  JHEP {\bf 0203} (2002) 046
  [arXiv:hep-ph/0203049].

\bibitem{Yao:2006px}
  W.~M.~Yao {\it et al.}  [Particle Data Group],
  J.\ Phys.\ G {\bf 33} (2006) 1.

\bibitem{Boyle:2008yd}
  P.~A.~Boyle {\it et al.},
  JHEP {\bf 0807}, 112 (2008)
  [arXiv:0804.3971 [hep-lat]].

\bibitem{Bijnens:1998fm}
  J.~Bijnens, G.~Colangelo and P.~Talavera,
  JHEP {\bf 9805} (1998) 014
  [arXiv:hep-ph/9805389].

\bibitem{Masjuan:2008fv}
  P.~Masjuan, S.~Peris and J.~J.~Sanz-Cillero,
  Phys.\ Rev.\ D {\bf78} (2008) 074028
  arXiv:0807.4893 [hep-ph].

\bibitem{Caprini:1999ws}
  I.~Caprini,
  Eur.\ Phys.\ J.\  C {\bf 13} (2000) 471
  [arXiv:hep-ph/9907227].

\bibitem{Ananthanarayan:2008mg}
  B.~Ananthanarayan and S.~Ramanan,
  arXiv:0811.0482 [hep-ph].

\bibitem{Omnes:1958hv}
  R.~Omnes,
  Nuovo Cim.\  {\bf 8} (1958) 316.

\bibitem{GimenoSegovia:2008sx}
  M.~Gimeno-Segovia and F.~J.~Llanes-Estrada,
  Eur.\ Phys.\ J.\  C {\bf 56} (2008) 557
  [arXiv:0805.4145 [hep-th]].

\bibitem{Leutwyler:2002hm}
  H.~Leutwyler,
  arXiv:hep-ph/0212324.

\bibitem{brodskyfarrar}
   S.~J.~Brodsky and G.~R.~Farrar,
  Phys.\ Rev.\  D {\bf 11} (1975) 1309.


\bibitem{guerrero}
  F.~Guerrero,
  Phys.\ Rev.\  D {\bf 57} (1998) 4136
  [arXiv:hep-ph/9801305];
  F.~Guerrero and A.~Pich,
  Phys.\ Lett.\  B {\bf 412} (1997) 382
  [arXiv:hep-ph/9707347].

\bibitem{Oller:2007xd}
  J.~A.~Oller and L.~Roca,
  Phys.\ Lett.\  B {\bf 651} (2007) 139
  [arXiv:0704.0039 [hep-ph]].

\bibitem{gasserleutwyler}
  J.~Gasser and H.~Leutwyler,
  Annals Phys.\  {\bf 158} (1984) 142.

\bibitem{weinberg}
  S.~Weinberg,
  Physica A {\bf 96} (1979) 327.

\bibitem{truong}
  A.~Dobado, M.~J.~Herrero and T.~N.~Truong,
  Phys.\ Lett.\  B {\bf 235} (1990) 134.

\bibitem{Truong:1988zp}
  T.~N.~Truong,
  Phys.\ Rev.\ Lett.\  {\bf 61} (1988) 2526.

\bibitem{Guerrero:1998ei}
  F.~Guerrero and J.~A.~Oller,
  Nucl.\ Phys.\  B {\bf 537} (1999) 459
  [Erratum-ibid.\  B {\bf 602} (2001) 641]
  [arXiv:hep-ph/9805334].

\bibitem{various}
  G.~Ecker, J.~Gasser, A.~Pich and E.~de Rafael,
  Nucl.\ Phys.\  B {\bf 321} (1989) 311;
  F.~J.~Llanes-Estrada and P.~De A. Bicudo,
  Phys.\ Rev.\  D {\bf 68} (2003) 094014
  [arXiv:hep-ph/0306146];
  T.~N.~Pham and T.~N.~Truong,
  Phys.\ Rev.\  D {\bf 31} (1985) 3027;
  F.~J.~Yndurain,
  arXiv:hep-ph/0212282;
  M.~R.~Pennington and J.~Portoles,
  Phys.\ Lett.\  B {\bf 344} (1995) 399
  [arXiv:hep-ph/9409426].

\bibitem{Dobado:1996ps}
  A.~Dobado and J.~R.~Pelaez,
  Phys.\ Rev.\  D {\bf 56} (1997) 3057
  [arXiv:hep-ph/9604416].


\bibitem{dgnmp}
  A.Dobado {\it et al.}, Effective Lagrangians for the Standard Model,
Springer-Verlag, Berlin Heidelberg 1997.


\bibitem{Bunton:2006va}
  T.~B.~Bunton, F.~J.~Jiang and B.~C.~Tiburzi,
  Phys.\ Rev.\  D {\bf 74} (2006) 034514
  [Erratum-ibid.\  D {\bf 74} (2006) 099902]
  [arXiv:hep-lat/0607001].


\bibitem{donoghue}
  J.~F.~Donoghue,
  arXiv:hep-ph/9607351.

\bibitem{Bruns:2004tj}
  P.~C.~Bruns and U.-G.~Mei\ss ner,
  Eur.\ Phys.\ J.\  C {\bf 40} (2005) 97
  [arXiv:hep-ph/0411223].




\end{thebibliography}
\end{document}